\documentclass{book}

\usepackage[paperwidth=17cm, paperheight=24cm, includeheadfoot,
            left=2cm, right=2cm, top=2cm, bottom=1.3cm, twoside]{geometry}

\usepackage%
 {
  algorithm2e, amsmath, amssymb, appendix, caption, cite, enumitem, fancyhdr,
  fancyvrb, graphicx, import, lineno, longtable, remreset, shorttoc, sidecap,
  url
 }

\newcommand{\tmpstring}{}
\newcommand{\settmpstring}[1]{\renewcommand{\tmpstring}{#1}}
\newcommand{\SourceCodeLines}[1]
 {%
  \settmpstring{{\ttfamily\bfseries\tiny\theFancyVerbLine}}
  \ifnum#1>9
    \settmpstring
     {\parbox[b]{7.5pt}{\ttfamily\bfseries\tiny\rightline\theFancyVerbLine}}
  \fi
  \ifnum#1>99
    \settmpstring
     {\parbox[b]{11.2pt}{\ttfamily\bfseries\tiny\rightline\theFancyVerbLine}}
  \fi
 }

\def\thepart{\Alph{part}}

\makeatletter
\@removefromreset{figure}{chapter}
\makeatother

\makeatletter
\renewcommand{\thefigure}{\@arabic\c@figure}
\makeatother

\makeatletter
\@removefromreset{table}{chapter}
\makeatother

\makeatletter
\renewcommand{\thetable}{\@arabic\c@table}
\makeatother

\makeatletter
\@removefromreset{equation}{chapter}
\makeatother

\makeatletter
\renewcommand{\theequation}{\@arabic\c@equation}
\makeatother

\usepackage{tocloft}
\usepackage{minitoc}

\cftsetpnumwidth{6.1mm}

\makeatletter
\renewcommand{\@tocrmarg}{4em}
\makeatother

\mtcsetformat{minitoc}{tocrightmargin}{2.55em plus 1fil}

\newcommand{\Author}{}
\newcommand{\AuthorLastName}[1]{\renewcommand{\Author}{#1}}

\def\Title#1{\chapter[\thepart\thelecture\ \protect\mbox{\protect%
\parbox[t]{110mm}{#1 \textit{(\Author)}}}\smallskip]{\Large #1}}

\newsavebox{\shortTitleBox}
\def\shortTitle#1{\savebox{\shortTitleBox}{#1 \textit{(\Author)}}}

\newcounter{lecture}[part]

\fancypagestyle{plain}
 {
  \renewcommand{\headrulewidth}{0pt}
  \fancyhead[OL]
   {
    \vspace{-8.5pt}
    \textit{Lecture given at the International Summer School
   \emph{Modern Computational Science}}\\
    \textit{(August 9-20, 2010, Oldenburg, Germany)}
  }
 }

 {
  \renewcommand{\headrulewidth}{0.4pt}
  \fancyhf{}
  \fancyhead[OR]{\rightmark}
  \fancyfoot[OR]{\thepage}
  \fancyfoot[EL]{\thepage}
  \fancyhead[EL]{\usebox{\shortTitleBox}}
 }

\newcommand{\SwitchToFancy}
 {%
  \pagestyle{fancy}
   {
    \renewcommand{\headrulewidth}{0.4pt}
    \fancyhf{}
    \fancyhead[OR]{\rightmark}
    \fancyfoot[OR]{\thepage}
    \fancyfoot[EL]{\thepage}
    \fancyhead[EL]{\usebox{\shortTitleBox}}
   }
 }

\makeatletter
\def\thebibliography#1
{
 \section*{References\@mkboth{References}{References}}
 \list{[\arabic{enumi}]}
  {
   \settowidth\labelwidth{[#1]}\leftmargin\labelwidth\advance\leftmargin
   \labelsep\usecounter{enumi}
  }
 \def\newblock{\hskip .11em plus .33em minus .07em}
 \sloppy\clubpenalty4000\widowpenalty4000\sfcode`\.=1000\relax
}
\makeatother

\makeatletter
\renewcommand{\@makefntext}[1]{\setlength{\parindent}{0pt}%
\begin{list}{}{\setlength{\labelwidth}{1.5em}%
\setlength{\leftmargin}{\labelwidth}%
\setlength{\labelsep}{3pt}\setlength{\itemsep}{0pt }%
\setlength{\parsep}{0pt}\setlength{\topsep}{0pt}%
\footnotesize}\item[\hfill\@makefnmark]#1%
\end{list}}
\makeatother

\begin{document}

\dominitoc

\faketableofcontents

\renewcommand{\cftsecfont}{\bfseries}
\renewcommand{\cftsecleader}{\bfseries\cftdotfill{\cftdotsep}}
\renewcommand{\cftsecpagefont}{\bfseries}

\setlength{\cftsubsecindent}{12.5mm}

\captionsetup{width=0.9\textwidth,font=small,labelfont=bf}


\stepcounter{lecture}
\setcounter{figure}{0}
\setcounter{equation}{0}
\setcounter{table}{0}


\AuthorLastName{Katzgraber}

\Title{Scientific Software Engineering in a Nutshell}

\shortTitle{Scientific Software Engineering}

\SwitchToFancy

\bigskip
\bigskip


\begin{raggedright}
  \itshape Helmut G.~Katzgraber\\
  \bigskip
  Department of Physics and Astronomy, Texas A\&M University\\
  College Station, Texas 77843-4242 USA\\
  \medskip
  Theoretische Physik, ETH Zurich\\
  CH-8093 Zurich, Switzerland\\
  \bigskip
  \bigskip
\end{raggedright}


\paragraph{Abstract.} Writing complex computer programs to study
scientific problems requires careful planning and an in-depth knowledge
of programming languages and tools. In this lecture the importance
of using the right tool for the right problem is emphasized. Common
tools to organize computer programs, as well as to debug and improve
them are discussed, followed by simple data reduction strategies and
visualization tools. Furthermore, some useful scientific libraries
such as boost, GSL, LEDA and numerical recipes are outlined.


\minitoc

\section{Introduction}
\label{sec:intro}

Taking on scientific problems using computers, while not completely new,
is still one of the branches of scientific research which is least
well documented---especially for beginners. There is no ``right''
or ``optimal'' way to develop software to study scientific problems,
because this is an extremely problem-dependent task. For example,
studying events in high-energy physics experiments requires thousands
of independently-working CPUs, simulations of hydrodynamic processes
in general require massively-parallel machines, whereas many of
every day's numerical problems we encounter can be solved on a simple
desktop workstation or laptop computer.

The goal of this tutorial is two-fold: First and foremost, to
convey the necessary tools and ``organizational skills'' to develop
(small-scale) computer programs for scientific simulations. The
tutorial focuses on different aspects one should pay attention to
before, during, and after the software development phase. Second,
emphasis is placed on the use of the right language for the right
problem. For example, it makes little sense to write a program in a
programming language such as C or FORTRAN to numerically integrate a
function, if software packages such as Mathematica or Matlab can do
this within one line of code.

Clearly, it is impossible to convey all the necessary information
in this brief tutorial. Thus emphasis is placed on the necessary
references where students can search for and find the necessary
information.  In this tutorial it is assumed that the reader is
familiar with a basic *nix environment (e.g., Linux, HP-UX, MacOS
X, \ldots) and has knowledge of at least one high-level programming
language (e.g., FORTRAN, C, C++, Java, \ldots). Unfortunately, it is
beyond the scope of this tutorial to introduce the *nix environment or
a high-level language. The reader is thus referred to a vast variety of
freely-available online tutorials, as well as the book ``{\em Learning
the Unix Operating System}'' by Peek {\em et al.}~in the O'Reilly
Bookshelf \cite{comment:oreilly}.  If necessary, examples are written
in the C programming language and shell commands are presented using
bash shell syntax \cite{comment:bash}.  General notation: Source code
and commands are typeset \texttt{in this font} and a *nix prompt in
a shell is represented by ``\texttt{>}.''

\section{General strategy}
\label{sec:strategy}

In general, writing a computer program to study some scientific problem
can be divided into different phases: planning, coding, debugging and
testing, data production, data analysis, documentation for posterity
and publication of the results. First, the general strategy for
software development is discussed, followed by some useful tools.

General rule: Never simply start writing a program for some
research project. Most of the time, programs written in an ad-hoc
way turn out to be ``Franken-programs'' that are hard to read,
difficult to adjust to different problems, and even slow. This is
particularly the case when more than one researcher is working on
a given software project. Therefore some {\em software engineering}
should be done. Below, the basic steps of software engineering within
a scientific context \cite{sommerville:89,hartmann:01} are outlined.
See, for example, Ref.~\cite{hartmann:01} for further details.

\subsection{Definition and outline of the problem}
\label{sec:define}

Probably the first task to accomplish is to decide which quantities
need to be computed using which algorithm/method to study a given
computational problem. Think also about other quantities which {\em
could} be measured (at small computational cost) that might be of
use for later projects and might not be necessarily relevant for the
current problem. Some points to consider:

\begin{itemize}

\item[$\Box$]{

Draw a flowchart (diagram) for the problem. Think about the input,
routines needed (and their order), as well as the output. Make sure
you are using the right algorithm and programming language to solve
the problem (check the literature and talk to your peers!).

}

\item[$\Box$]{

Make a list of necessary input parameters for the
simulation. Parameters that change often or alter the functionality
of the code can be passed as run-time options (e.g., space dimension,
seed of the random number generator, data compression option, etc.).
It is recommended to use a parameter file for all other parameters.
Keeping such a parameter file together with the produced data is of
paramount importance to ensure data provenance (discussed later).

}

\item[$\Box$]{

Because the cost of storage has decreased considerably in the last
few years, it is recommended to store as much simulation information
as possible. Thus, make a list of potential observables (quantities
to be measured), and, as mentioned before, any other observables
that might be useful elsewhere, and decide to what level data
have to be stored. For example, in a Monte Carlo simulation one can
store the final thermal average of some quantity, one can store the
time evolution (logarithmically-spaced) of the same quantity, or a
configuration snapshot of every single step in the simulation. Clearly,
the storage requirements increase considerably.

}

\item[$\Box$]{

Identify {\em objects} and {\em data structures} in your problem. This
shall enhance the modularity of the program. For example, position
and velocity of a particle can be combined into one structure dubbed
``particle'' thus allowing for an easy change of coordinate systems
or space dimensions. Use a bottom-up approach to programming: Define
essential routines and code them up first. Add the ``skeleton''
of the simulation at the end.

}

\item[$\Box$]{

Think back: Do you currently have programs or libraries that would
be of use for this project and can be included? Think ahead: Will
this program be used in the future for other projects? If so, ensure
that extensions can be trivially accomplished and do not require a
complete re-writing of the software.

}

\end{itemize}

\noindent Once the planning phase is complete, it is important to
decide which language to use for a given project.

\subsection{Selecting a programming language}
\label{sec:language}

In general, one can roughly classify computer languages for
scientific applications into three (overlapping) categories: high-level
programming languages (such as C, C++, FORTRAN, Java, etc.), scripting
languages (e.g., Perl, Python, shell, R) and ``symbolic languages''
(e.g., Mathematica, Matlab, Maple, to name a few). It is of paramount
importance to select the right language for a given problem. Keep
the following points in mind when selecting a programming language:

\begin{itemize}

\item[$\Box$]{

What is the numerical effort in CPU hours you expect the project to
take? If this number is very large, then a fast high-level language
should be used unless the complexity of the problem requires the use
of a symbolic language.  Furthermore, large projects can be sped up by
using parallel programming techniques on large clusters. In general,
these Message Passing (MPI) libraries \cite{comment:mpi} only exist
for high-level languages and some software packages such as Matlab.

}

\item[$\Box$]{

Does the problem require a real-time graphical display of data or
cross-platform compatibility? In that case Java (by Sun Microsystems)
might be the best choice. Note that Java is a (relatively slow)
object-oriented language.

}

\item[$\Box$]{

Does the problem require parsing and analyzing large data sets? In
this case a scripting language such as Perl or Python might be most
useful because of the built-in Regular Expression parsing capabilities
\cite{comment:regexp}. Furthermore, in the case of Perl, input-output
tasks are trivially accomplished and so the combination of different
data sets or files for parsing is done very efficiently. Note that
these languages do not require compilation and thus are very portable
between operating systems and different hardware architectures.

}

\item[$\Box$]{

Does the problem require symbolic manipulations of functions or,
e.g., complex numerical integrations of functions?  In this case
a symbolic language such as included in Matlab or Mathematica
might be the optimal choice to tackle the problem. For example,
Mathematica allows for axiomatic definitions of operators. This
in turn simplifies considerably the computation of, e.g.,  Feynman
diagrams for large-order expansions.

}

\item[$\Box$]{

Does the problem use pre-defined software packages that require being
called with specific parameter combinations?  In this case it might
be most useful to ``wrap'' the used software in a shell (or Perl)
script to automatize the (often tedious) tasks involved in calling
the program. Furthermore, multiple analysis steps can be wrapped in
one script thus reducing effort and ensuring a minimal error rate.

}

\end{itemize}

\noindent Overall one has to consider the {\em total} amount of time
spent on a project, including the software development. For example,
why write a C program to integrate a function if Mathematica has a
(very good) built-in integration routine? This could save considerable
time in the solution process of a problem.

A typical example of how to split up tasks between programming
languages is illustrated with a Monte Carlo simulation of a spin
system. A good approach would be to write the main routine generating
the data using C or C++. To run the program written in a high-level
language on many workstations, one would use a shell script to
farm out and monitor the executables. To analyze the data---such as
for example computing thermal averages over measurements and error
bars---one could use Perl scripts for post-processing.

\subsection{Writing the program}
\label{sec:program}

Using a consistent notation for variables, as well as a clear
pre-defined style of programming ensures the portability and---most
importantly---readability of the code after extended periods of
time. General considerations:

\begin{itemize}

\item[$\Box$]{

Use a modular programming approach: Split the code into several
modules that can be stored in independent source code files. This
has the following advantages: It allows you to easily use these
modules for other programs. For example, a C-routine to compute the
mean of an array of numbers can be stored in a file \texttt{mean.c}
which then can be used for other coding projects by simply copying
the file or loading a library (discussed later). If you have a large
software project, changes to individual modules will only require
recompiling of these object files. Finally, if multiple researchers
are working on one project, having only one source file prevents
people from working on the same file in parallel.

}

\item[$\Box$]{

To keep the program logically structured, separate generic data
structures and algorithms in separate files, such as~\texttt{.c}
and~\texttt{.h} files in C.

}

\item[$\Box$]{

Give meaningful names to variables, routines, etc. For example, a
subroutine \texttt{e(int *s)} carries little to no information for
the reader. Instead, naming the routine and arguments in the following
way \texttt{energy(int *spins)} is far more informative. Be sure you
use a {\em consistent} notation. Example:

\begin{verbatim}
 #defines USING_ALL_CAPS
 variables in smallCamelCase
 local variables _withUnderscore
 classes in CapitalizedCamelCase
 functions small_with_underscores( )
\end{verbatim}

\noindent If you choose to use short names for variables, be sure to
document their meaning.

}

\item[$\Box$]{

Try to use proper indentation when writing routines as this increases
readability considerably. Good style:

\SourceCodeLines{9}
\begin{Verbatim}[fontsize=\small]
 #include <stdio.h>

 int main(){

    int counter;
   
    for(counter = 1; counter < 100; counter++){
        printf("Asterix beats the Romans up. Ha!\n");
    }

    return(0);
 }
\end{Verbatim}

Bad style:

\SourceCodeLines{9}
\begin{Verbatim}[fontsize=\small]
 #include <stdio.h>
 int main()
 {int i;
 for(i = 1; i < 100; i++) printf("Asterix beats the Romans up. Ha!\n");
 return(0);}
\end{Verbatim}

Adding extra space {\em always} makes the code easier to read,
{\em but} do not go too crazy as then it might be too hard to read,
too. A good measure is a four-character indentation as shown in the
example above. Most editors do this automatically.

}

\item[$\Box$]{

Avoid logical jumps (goto) in the program. Debugging will be a
nightmare otherwise.

}

\item[$\Box$]{

Be careful when using global variables. It makes sense to define some
global variables, but, in general, it is {\em highly} recommended
to pass variables via a ``global-variables'' structure to avoid
programming mistakes. Quantities that do not change in the simulation
and which are needed by many routines, could, in principle, be defined
globally (e.g., the space dimension or the number of particles).

}

\end{itemize}

\noindent One of the most important steps in writing software is to
properly document it. Always write at the beginning of the routine a
short introduction about the routine and what it does. You can also
add a brief history with different versions and revisions. List
parameters as well as their meaning. Nice examples are shown
in Ref.~\cite{hartmann:01}. Within the code, list the meaning of
variables, comment on what certain routines for a code block do and
always remember that a few years down the road it might not be quite
clear what you were doing. Furthermore, provide external documentation
for other users (generally in form of a README file).

Finally, you might wonder in which environment/editor the code should
be written in. While some developers rather use IDEs (integrated
development environments) such as {\em eclipse} \cite{comment:eclipse}
or Apple Inc.'s {\em Xcode} \cite{comment:xcode}, others prefer simple
*nix editors such as {\em emacs} or the {\em vi} editor. This again
depends strongly on the scope and complexity of the project, as well
as personal preference and habits.

\subsection{Help! I am lost!}
\label{sec:help}

Do not despair. Almost all programs have built-in help. In general,
when these programs are started from a command line, you can obtain
a bare-bones help by using the following {\em flags}:

\begin{verbatim}
 > program -h
\end{verbatim}

\noindent or

\begin{verbatim}
 > program --help
\end{verbatim}

\noindent Always try both since sometimes one is more verbose than
the other. In addition to the built-in help, any *nix operating system has
a built-in system of manual pages. The system can be started in a shell by
issuing the command \texttt{man}. For example, to find out more about the
\texttt{man} command itself, simply issue:

\begin{verbatim}
 > man man 
\end{verbatim}

\noindent Scrolling is accomplished with the arrow keys, searching
within the man-page for  a text string ``\texttt{string}''can be done
with ``\texttt{/string}'' and quitting the system is done by pressing
``\texttt{q}''. If you are not quite sure what you are searching
for, you can issue the wild card search command, here for the string
\texttt{cron}

\begin{verbatim}
 > man -k cron
\end{verbatim}

\noindent which produces:

\begin{verbatim}
 cron(8)                  - daemon to execute scheduled commands
 crontab(1)               - maintain crontab files for users
 crontab(5)               - tables for driving cron
\end{verbatim}

\noindent The output has a small description. Note that there are two
man-pages for ``crontab.'' To access page 5, simply issue

\begin{verbatim}
 > man crontab -S5
\end{verbatim}

\noindent Certain core components of the standard C library are also
documented via man-pages. Therefore, when programming in C, information
about any built-in function can be obtained.  ``\texttt{man sin}''
returns information on how to call the sine function, which arguments
the function call has, as well as which header files to load (in this
case \texttt{math.h}).

Finally, there is always the internet and Wikipedia as sources of
information. Be always aware of what the sources are since there is
no guarantee that the information retrieved is correct.

\section{Provenance}
\label{sec:provenance}

In the previous section the importance of readability of the code
as well as documenting the code has been emphasized. One of the
most important reasons for careful programming is code and data
{\em provenance}. In the Oxford dictionary {\em provenance} is
defined as ``a record of ownership of a work of art or an antique,
used as a guide to authenticity or quality.'' While this definition
is explicitly given for antiques and pieces of art, the concept of
provenance has become of increasing importance in the last few years
within the context of computational sciences. As computer programs
become increasingly complex and data sets grow exponentially with
time, it is of paramount importance to document the {\em origins}
of the code and data, the different {\em revisions} and versions,
as well as the {\em usage} of code and data analysis tools.

Although there are no world-wide standards on how to properly
store computer codes or data files---each of which depend highly
on the application or problem---, there have been some incentives
to standardize at least data files to ensure that results published
in scientific journals can be reconstructed in the future.  In this
section we differentiate slightly between source code provenance
and data provenance. Probably the easiest way to ensure source code
provenance is to use a version control system which keeps track of
changes made to software.

\subsection{Version control with subversion}

In addition to source code provenance, version control systems (VCSs)
have the advantage of allowing one to catalog in an easy way older
versions of a given document (text file, program, etc.) for retrieval.
This is particularly useful when more than one person is working
on a program since then individual portions of the code can be {\em
checked out}, ensuring that two persons are not editing the same file
at the same time. Once editing is complete, the code is {\em committed}
to the VCS for others to access.

{\em Subversion} \cite{comment:subversion} is a VCS that can manage
any project situated on a directory containing any arbitrary files
on your hard disk. Note that the system does not store copies of
all revisions made, it merely stores the {\em changes} between the
different versions. By default, subversion does not {\em lock} the
files, i.e., more than one user can edit a given file. If this is
undesired, file locking can be turned on.

To create a repository located in the directory
``\texttt{/home/user/repository}'' issue the command

\begin{verbatim}
 > svnadmin create /home/user/repository
\end{verbatim}

\noindent All operations performed on the repository are done with the
command \texttt{svn}. The general syntax is \texttt{svn <subcommand>
[options] [args]}. More help can be obtained by simply typing
\texttt{svn help}.  If you already have some files in a directory
called ``\texttt{lala},'' you can {\em import} them to the repository
by specifying a destination directory as well as a {\em message}
via the \texttt{-m} option:

\begin{Verbatim}
 > svn import lala/ file:///home/user/repository/lala -m "first import"
\end{Verbatim}

\noindent Note that we use a universal resource locator (URL) syntax
for the placement of the file, in this case ``\texttt{file://}'' since
it resides on the local disk. This means that networked projects
can be accessed via ``\texttt{http://}'' or ``\texttt{ftp://}''
(check the documentation). To view the files in the repository and within
the created directory ``\texttt{lala}'' type:

\begin{verbatim}
 > svn list file:///home/user/repository/
  
 > svn list file:///home/user/repository/lala 
\end{verbatim}

\noindent Before you start editing any files be
sure to check out the files with 

\begin{verbatim}
 > svn checkout file:///home/user/repository/lala
\end{verbatim}

\noindent or any specific file within the repository by specifically
specifying it. You can check the status of which files have been
edited by using the command

\begin{verbatim}
 > svn status
\end{verbatim}

\noindent Once you have completed the necessary edits of the files,
you can {\em commit} your changes and upload them to the repository via

\begin{verbatim}
 > svn commit -m "did some lousy changes"
\end{verbatim}

\noindent Note that again a message is passed, documenting the changes
made. {\em While} you have a project checked out you can always see
the changes you have made with

\begin{verbatim}
 > svn diff
\end{verbatim}

\noindent The output is in the Unix diff format which allows you
to patch a file with the provided text (see \texttt{man diff}
and \texttt{man patch}). If other authors performed changes to the
repository's files, you can obtain the latest revision by using the
command ``\texttt{svn update}.'' Potential conflicts can be resolved
using the ``\texttt{resolve}'' option. Finally, the revision history
can be generated using ``\texttt{svn log}.'' There are many other
options and the reader is referred to the documentation as well as
Ref.~\cite{collins-sussman:08}.

Finally, there are also distributed source control management
tools. For example, Mercurial \cite{comment:mercurial} ensures that
each developer has a local copy of the source code and the entire
development history. While subversion relies on a central server to
store the revisions, code development on Mercurial is independent of
a central server and therefore also network independent. Note that
Mercurial is written in Python and therefore platform independent,
as well as open source.

\subsection{Data provenance}

In addition to the need for proper documentation of source codes, it is
imperative to have consistent data formats for numerical simulations
that are well documented with external READMEs. Currently, it is
unclear what the best format for data storage is. Large and complex
data sets are generally binary compressed and stored using {\em hdf5}
\cite{comment:hdf5}. For now, we assume that this is not the case
and that storing data in plaintext files is appropriate. The optimal
approach to storing data is using XML tags (see books on the extended
markup language for details, e.g., Ref.~\cite{means:01}) with the
disadvantage that data files are extremely large because the XML tags
require a nonnegligible number of bytes.  In addition, data stored in
XML format are generally hard to read and require parsers to produce
sets which are easily understood. Therefore, a good compromise solution
is to generate a well-documented tabular data structure.  A typical
bad example on how data are stored in tabular manner is given below:

\begin{verbatim}
 2.0000000000e+00 -1.7430877686e+00 
 2.2000000000e+00 -1.5684322497e+00
 2.4000000000e+00 -1.3377984675e+00
 2.6000000000e+00 -1.1194154460e+00
\end{verbatim}

\noindent There is no description of the data, there are no tags of any
kind, there is no information on how or when it has been generated.
Therefore, it is impossible for a third-party to understand the
results. In this case, it is the temperature-dependent internal
energy for a two-dimensional Ising \cite{ising:25,yeomans:92} model
with $8^2$ spins computed using Monte Carlo methods. There is no way
to check in hindsight if the data are converged since only averages
at a given number of Monte Carlo steps have been stored. To be able
to ensure that the data are converged, a whole new simulation would
have to be performed.

By adding a (parsable) header file to the data set, it is easy to know
what is stored in the file. This is shown in the example below (excerpt
only) which has such a header with the simulation parameters, as well
as the process ID in case one needs to kill a specific simulation.
Furthermore, the temperatures are listed (for parsing and analysis
convenience) as well as the average energy data as a function of
(logarithmic) Monte Carlo time to be able to check if the data are
converged. Time-dependent blocks for different temperatures are
separated by empty lines.

\begin{verbatim}
 # 2D ferromagnetic Ising code. h. katzgraber 12/06/2009 (v4.20)
 # simulated using simple Monte Carlo
 #
 # system size (number of spins)         L  = 8 ( N = 64 )
 # space dimension                       D  = 2
 # max exponents for sweeps (EXP_MAX)       = 12
 # maximum equilibration/measurement time   = 4096
 # initial seed for this run                = 70893
 # process id (PID) for this run            = 89993
 # temperature set:
 #
 # | 2.0000
 # | 2.2000
 # | 2.4000
 # | 2.6000
 #
 # T    MCS     <energy>
 2.0000 2      -1.5312500000e+00
 2.0000 4      -1.8281250000e+00
 2.0000 8      -1.7812500000e+00
 2.0000 16     -1.7734375000e+00
 2.0000 32     -1.7851562500e+00
 2.0000 64     -1.7900390625e+00
 2.0000 128    -1.7685546875e+00
 2.0000 256    -1.7634277344e+00
 2.0000 512    -1.7650146484e+00
 2.0000 1024   -1.7451782227e+00
 2.0000 2048   -1.7450561523e+00
 2.0000 4096   -1.7430877686e+00

 2.200 2 ...
\end{verbatim}

\noindent Note that the data file also includes a version of the
used program. If the source code used for the project is under
continuous development, it is very important to know which version of
the program was used to generate a given data set.  Clearly, there
is no best recipe since the way data are stored depends highly on
the problem studied. Nevertheless it is crucial to put some thought
into how much of the data and how the data are stored for posterity
and provenance. Furthermore, always document the data, software and
analysis tools.

\section{Compiling, debugging, profiling, testing}
\label{sec:comptest}

Once the code is written, it will likely not compile. That is the
best case scenario since compilers, in general, deliver useful error
messages. If the code compiles, and crashes during the first run,
this is not too bad either since debuggers can assist in finding the
problem. If the code compiles and the program exits normally, but the
data you produce do not seem to be correct, be very afraid since these
are the hardest bugs to find.  In this section compilers, options,
and basic compilation procedures are discussed, followed by common
debugging tools. In addition, some suggested approaches for finding
hidden problems {\em if} the code compiles and runs are presented.

\subsection{Compilers and options}

There are many different compilers for many different languages. For
this reason in this section we focus on the GNU C compiler
\texttt{gcc}. There are commercial compilers (Intel, Portland Group,
Numerical Algorithms Group) that, in general, produce considerably
faster executables than the GNU C compiler. Hence, if you have
access to these (e.g., on a large computer cluster) you should try
to use them.

After the code is written, it is compiled with a compiler.  The
compilation process encompasses several stages, starting with the
preprocessor which resolves \texttt{\#define}, \texttt{\#include}
and \texttt{\#if} directives. In this case \texttt{gcc} actually
invokes the preprocessor \texttt{cpp} to do the preprocessing. Once
the preprocessing is completed, the compiler produces machine-readable
assembly language from the input files (usually this step produces no
visible output, but, if requested, it will produce assembler files
with a ``\texttt{.s}'' ending). Within \texttt{gcc} the assembler
\texttt{as} takes the produced {\em assembler} code and generates
object files (with the ending ``\texttt{.o}''). In the final stage,
the objects \texttt{.o} are placed in their proper place of the
executable. Library functions might be included at this stage as
well.  \texttt{gcc} invokes internally the {\em linker} \texttt{ld}
for this task.  Note that, in general, all these steps are performed
automatically by the compiler. You {\em do} have the option to stop the
compiler at any stage and invoke the individual steps manually. Because
this would be beyond the scope of this lecture, the reader is referred
to Ref.~\cite{loukides:97}.

\paragraph{Basic compiler use} The most basic way to call the compiler
is in the following way:

\begin{verbatim}
 > gcc main.c lala.c momo.c
\end{verbatim}

\noindent This produces an executable \texttt{a.out} which then can
be executed. Programmers rarely build everything at once. Usually they
attempt to compile individual parts of the program to check for errors
before the whole program is compiled. Thus, conversely, we can call

\begin{verbatim}
 > gcc -c main.c 
 > gcc -c lala.c 
 > gcc -c momo.c
 > gcc main.o lala.o momo.o
\end{verbatim}

\noindent to obtain the same result. The ``\texttt{-c}'' options means
``compile but do not link.'' At the end \texttt{gcc} is invoked with
only object files to henceforth link them and produce an executable.

If library functions are to be linked into the executable, we need
to call these with the ``\texttt{-l}'' option. In this case, if the
library is called for example \texttt{libmy.a} and it is needed for
the object compilation of \texttt{main.o} we need to issue

\begin{verbatim}
 > gcc -c main.c -lmy
\end{verbatim}

\noindent Libraries are discussed in more detail later.  If we
want to include an include file or library which is in a nonstandard
directory, we need to tell the compiler where to find these. For
example:

\begin{verbatim}
 > gcc -c main.c -I/home/user/myinclude -L/home/user/mylibs -lmy
\end{verbatim}

\paragraph{Summary of some compiler options}

There is a whole zoo of compiler options that influence the behavior
and speed of your executable, as well as show in a verbose manner
warnings and comments during the compilation of the program. There is
a core subset of compiler options which not only are very important,
but also more or less carry over to different programming languages
and compilers. In what follows the most important compiler options
for the GNU C compiler (\texttt{gcc}) are listed. For a complete list,
\texttt{man gcc} as well as Ref.~\cite{loukides:97} can be used.

\begin{itemize}

\item[]{\hspace*{-1em}\texttt{-v}\newline Prints the compiler version
and details about the configuration.}

\item[]{\hspace*{-1em}\texttt{-o}\newline Sets the name of the
output file.}

\item[]{\hspace*{-1em}\texttt{-c}\newline Compile only, do not
link. Produces object files.}

\item[]{\hspace*{-1em}\texttt{-lany}\newline Tells the compiler
to link the objects to a library called \texttt{libany.a}.}

\item[]{\hspace*{-1em}\texttt{-static}\newline Link only to static
libraries. This is sometimes necessary for executable portability to
other machines where the libraries are not installed.}

\item[]{\hspace*{-1em}\texttt{-Wall}\newline Turn on (almost) all
compiler warnings. The compiler warns if there are potential problems
in the code.}

\item[]{\hspace*{-1em}\texttt{-g}\newline Turn debugging on and
generate an expanded symbol table.  Only code compiled with this flag
produces meaningful output in common debugging programs such as
\texttt{gdb} (see Sec.~\ref{sec:debuggers}). The symbol table can,
in principle, later be removed with the Unix \texttt{strip} command.}

\item[]{\hspace*{-1em}\texttt{-pg}\newline Link the program for
profiling with \texttt{gprof}. This produces execution statistics
and timing information.}

\item[]{\hspace*{-1em}\texttt{-D}\newline Defines macros (see compiler
documentation). Useful for debugging or making architecture-dependent
code.}

\item[]{\hspace*{-1em}\texttt{-L} and \texttt{-I}\newline Tells the
compiler where to find custom library and header files (explained
above), respectively.}

\item[]{\hspace*{-1em}\texttt{-O}{\em n}\newline Optimizes the
code to level $n$. $n = 0$ means no optimization, whereas $n =
1$ means that the compiler tries to reduce the size of the code
as well as execution time. Compilation is slower and requires more
memory.  For production, it is optimal to compile with $n = 2$, i.e.,
\texttt{-O2}.  In this case the code is considerably faster than when
no optimization is used. $n = 3$ should be thoroughly tested before
used for production as ``experimental'' optimization features are
turned on. If possible, avoid. There are further custom optimization
options such as \texttt{-ffast-math}, \texttt{-finline-functions},
\texttt{-funroll-loops}, \ldots, as well as architecture-specific
instructions. Details can be found in the compiler documentation.}

\end{itemize}

\noindent In general, when it comes to speeding up an executable,
a careful choice of compiler options can yield great speed
improvements. Loop unrolling and inlining, combined with \texttt{-O2}
give the greatest performance boosts.

\subsection{Make}

The \texttt{make} facility \cite{oram:91} is one of Unix's most
useful tools. Unfortunately, it is one of the least used by beginning
programmers. Putting it in simple terms, \texttt{make} is a programming
language to automatize large compilations. Not only does \texttt{make}
handle automatic compilation of computer programs, the compilation of
large \LaTeX~documents whilst including tables of contents as well as
bibliographies can be simplified enormously by using \texttt{make}.
\texttt{make} checks if certain files have changed and, after
checking pre-defined dependencies in a configuration file called {\em
Makefile}, will {\em only} compile the necessary files and perform the
final linking. Not only does this considerably speed up compilation
time for large projects, it also reduces possible errors when linking
to old object files.

The {\em Makefile} contains the information about the different {\em
dependencies} between the source files as well as the necessary {\em
commands} to deal with these.  In general, the goal is to complete
a task called a {\em target}. A pair of dependencies and commands is
referred to as a {\em rule}. General syntax:

\begin{verbatim}
 target: sources
 <TAB>   commands
\end{verbatim}

\noindent The first line contains the dependencies for a given
target, the second one the commands to perform. Note that the
command line {\em must} always begin with a tabulator shift
(\texttt{<TAB>}). In all following examples this \texttt{<TAB>}
is not shown explicitly. Example:

\begin{verbatim}
 all: main.o lala.o
         gcc -o runme main.o lala.o

 main.o: main.c program.h
         gcc -c main.c

 lala.o: lala.c
         gcc -c lala.c
\end{verbatim}

\noindent In the previous example the file \texttt{main.o} has to be
compiled if either \texttt{main.c} or \texttt{program.h} have changed.
On the command line you run \texttt{make} via

\begin{verbatim}
 > make 
\end{verbatim}

\noindent If you want the program to be built, you would do
\texttt{make all}, and if you just want to recompile \texttt{lala.c},
you would do \texttt{make lala.o} Whenever the date of the source
files is newer than the date of the targets, \texttt{make} will
recompile these. In the following case (which is only practical for
small projects), all files are compiled every time:

\begin{verbatim}
 main:
         gcc -o runme main.c lala.c
\end{verbatim}

\noindent \texttt{make} checks recursively the dependencies in the
Makefile and executes the necessary commands.  \texttt{make} also
allows variable definitions ({\em macros}).
For example, one could define

\begin{verbatim}
 DEBUG = -g -Wall -ansi
 SRC   = *.c
\end{verbatim}

\noindent Then, these macros could be used later in the Makefile
allowing for easy global replacements of, e.g., the compiler or
debugging flags:

\begin{verbatim}
 all:
         gcc -o runme $(DEBUG) $(SRC)
\end{verbatim}

\noindent The behavior of \texttt{make} can also be altered with
run-time flags. Here is a list of the most convenient:

\begin{itemize}

\item[]{\hspace*{-1em}\texttt{-f} {\em filename}\newline Normally,
\texttt{make} looks for a file called {\em Makefile}. With this flag
the name can be changed to any {\em filename}.}

\item[]{\hspace*{-1em}\texttt{-n}\newline Do not execute any command,
simply list what would be executed (debugging).}

\item[]{\hspace*{-1em}\texttt{-i}\newline Normally, \texttt{make}
terminates if it encounters an error. This options forces \texttt{make}
to continue with the other remaining tasks.}

\item[]{\hspace*{-1em}\texttt{-j} {\em n}\newline Run {\em n} commands
at once (useful on multiprocessor machines).}

\end{itemize}

\noindent There are far more options and complexity to the
\texttt{make} command (see the manpage and Ref.~\cite{oram:91}).
Note also that line carriages can be accomplished with a
``\texttt{$\backslash$}''.

\paragraph{The one-size-fits-(almost)-all Makefile} For most
applications, the following Makefile---while requiring recompilation
of the whole source---provides the basic functionality and efficiency
needed.  Keep in mind: commands start with a \texttt{<TAB>}:

\SourceCodeLines{99}
\begin{Verbatim}[fontsize=\small]
 # check the hostname (hashed lines are comments)
 HN := $(shell /bin/hostname -s)

 # set executable name
 EXECNAME    = runme 

 # default compiler settings
 CC          =  gcc
 OPT         = -O2 
 DEBUG       = -g -Wall -ansi
 LDFLAGS     = -lm
 INCLUDE     = -I/usr/include -I../include

 # on moo.tamu.edu we use the intel compiler
 ifeq ($(HN),moo)
     CC      = icc
     OPT     = -O2 -static 
 endif

 SRC         = *.c
 OBJS        = $*(SRC).o

 # generic compilation for data production
 $(EXECNAME):
         $(CC) $(OPT) $(SRC) $(INCLUDE) -o $(EXECNAME) $(LDFLAGS) 
         /bin/rm -rf *.o

 # debugging compilation
 db:
         $(CC) $(DEBUG) $(SRC) $(INCLUDE) -o $(EXECNAME).db $(LDFLAGS) 
         /bin/rm -rf *.o

 # clean up
 clean:
         rm -rf *.o core *~ $(EXECNAME) $(EXECNAME).db
\end{Verbatim}

\noindent The aforementioned Makefile shows some crucial ingredients
needed when compiling simple projects on different
machines. Because some compilers are only installed on certain
machines, in line 2 we check the hostname of the machine using a shell
call. In line 5 we set a variable with the name of the executable and
in lines 8 -- 12 we define variables for the compiler, libraries,
optimization, etc.  From lines 15 -- 18 we check for a specific
hostname with an \texttt{ifeq} statement. If the code is compiled
on \texttt{moo} then we change \texttt{CC} from \texttt{gcc} to
\texttt{icc}. Note that we also perform a static compilation to
ensure that all libraries are contained in the final executable.
For simplicity, all source files are included using a wild card in line
20 and in line 21 the object files are defined by a replacement of the
file ending from \texttt{.c} to \texttt{.o}.  Production compilation
is accomplished by calling \texttt{make} or \texttt{make runme} in
lines 24 -- 26; the debugging compilation is called in lines 28 --
31. Note that after compilation the object files are deleted since
we want to ensure that the latest objects are included every time the
code is compiled. In lines 24 -- 25 we add a directive to ``clean up''
that deletes binaries, core files and object files.  Again, this is
not practical if the compilation takes several minutes. But if your
code compiles in 5 -- 10 seconds, this is possibly the easiest and
most versatile Makefile to use.

\subsection{Debuggers}
\label{sec:debuggers}

In this section we describe only the GNU \texttt{gdb} debugger. There
are graphical front-ends (such as \texttt{ddd}) for this debugger, but
these are not discussed here for the sake of brevity.  \texttt{gdb}
lets you run a C or C++ program, stop execution within the program,
examine and change variables during execution, call functions, and
trace how the program executes. Only the basic operation is presented,
more information can be found online at the Free Software Foundation's
{\em Debugging with GDB}, in Ref.~\cite{loukides:97}, the manpage,
as well as within the program by issuing the command ``\texttt{help}.''

For a program to be debuggable, one needs to use the ``\texttt{-g}''
compilation flag. Note that this makes the program very slow,
i.e., do not use this flag for production runs. Suppose a program
\texttt{runme.db} has been compiled. The debugger is invoked via

\begin{verbatim}
 > gdb ./runme.db
\end{verbatim}

\noindent The source code can be listed within the debugger with the
\texttt{list} command. To run the program within the debugger, simply
issue the \texttt{run} command. If your program requires arguments,
simply list them: \texttt{run -p params.in}.  To have the execution
stop at a specific line, one can use the \texttt{break} command,
i.e., \texttt{break 42} stops execution at line 42 of the source code
(note that the break-point needs to be set before the command is run).
You can \texttt{print} the contents of a variable or array, for example

\begin{verbatim}
 > (gdb) print array[10]
   $2 = 72
\end{verbatim}

\noindent whenever execution is stopped. Similarly, you can also
\texttt{set} the values of variables or arrays:

\begin{verbatim}
 > (gdb) set array[10] = 43
\end{verbatim}

\noindent To continue the execution of the program step-by-step,
use the \texttt{next} command; if you want the program to simply
execute normally (until it encounters the next break-point) use the
\texttt{cont} command. Note that \texttt{next} executes an entire
function if it encounters a call, while \texttt{step} enters the
function and keeps going one statement at a time. This allows the
programmer to fully control each step of the program to ensure that
things are running according to plan.

Because the breakpoints are numbered, during the execution conditions
can be placed with the \texttt{condition} command. It allows one
to stop execution if a condition is met -- something which is very
useful when testing.  Individual functions in the program can be
called with the \texttt{call} command and \texttt{info} delivers
{\em any} information about breakpoints, registers, variables, etc.
Finally, all aforementioned commands can be abbreviated by simply
typing the first letter of the command only. For example \texttt{n}
has the same functionality as \texttt{next}. Further details about
the use of \texttt{gdb} can be found in Ref.~\cite{loukides:97}
and in the documentation.

\subsection{Memory debugging with valgrind}

If memory allocation is not done carefully, memory segments outside
a data structure could be addressed. In this case the program exits
abnormally with a \texttt{Segmentation fault} dumping a {\em core}.
In addition, forgetting to free memory segments (also known as
{\em memory leaks}) could quickly use up all the available RAM on
a computer and crash it. Because memory allocation errors often do
not cause the program to crash immediately at the point where the
error happened, these types of errors are difficult to find. The
reason for this seemingly erratic behavior is that we cannot directly
influence memory management of the operating system. In some cases,
writing data out of bounds might overwrite other variables causing
a crash of the program, in others not.  This is the reason why you
should {\em always} test a program with memory checkers even though
things might seem to be running correctly.

To successfully trace memory errors, memory checkers such as valgrind
\cite{comment:valgrind} are very useful. Valgrind is freely available
online. To debug a program \texttt{runme.db} simply type on the
command line

\begin{verbatim}
 > valgrind ./runme.db
\end{verbatim}

\noindent Again, to be able to determine where in the source code the
bug is located, it is important to compile with the \texttt{-g} flag.
If you use the ``\texttt{--db-attach=yes}'' flag, valgrind starts
the debugger for you to analyze the program in more depth. It is
recommended to use valgrind in verbose mode (use the \texttt{-v}
flag) to obtain detailed information about the different memory
problems the program finds.  To see a sample output of valgrind,
run the memory checker on the *nix program of your choice, e.g.,
\texttt{valgrind date}:

\newpage

\small
\begin{verbatim}
 > valgrind date
 ==6147== Memcheck, a memory error detector.
 ... further copyright output
 ==6147== 
 Tue Apr 14 20:51:27 CEST 2009
 ==6147== 
 ==6147== ERROR SUMMARY: 0 errors from 0 contexts (suppressed: 15 from 1)
 ==6147== malloc/free: in use at exit: 0 bytes in 0 blocks.
 ==6147== malloc/free: 9 allocs, 9 frees, 1,102 bytes allocated.
 ==6147== For counts of detected errors, rerun with: -v
 ==6147== All heap blocks were freed -- no leaks are possible.
\end{verbatim}
\normalsize

\noindent The number at the beginning of each line is the process
ID and can be ignored. For the particular case of \texttt{date}
no errors were detected. If errors are present, the output mentions
mainly the error count:

\small
\begin{verbatim}
 > valgrind ./runme.db
 ==6745== Memcheck, a memory error detector.
 ... further copyright output
 ==6745== 
 ==6745== ERROR SUMMARY: 27 errors from 2 contexts (suppressed: 0 from 0)
 ==6745== malloc/free: in use at exit: 0 bytes in 0 blocks.
 ==6745== malloc/free: 0 allocs, 0 frees, 0 bytes allocated.
 ==6745== For counts of detected errors, rerun with: -v
 ==6745== All heap blocks were freed -- no leaks are possible.
\end{verbatim}
\normalsize

\noindent Re-running valgrind with the \texttt{-v} option (considerable
output) shows exactly where the problems are located.  More information
on valgrind's options can be obtained by using \texttt{valgrind -h}.

\subsection{Testing, testing, testing, \ldots}

Once the code compiles and runs, it still does not mean that it is
producing the correct results. At this stage, if there is suspicion
of a problem, different checking mechanisms can be used.

In C programs, one can use \texttt{\#if} directives to have the
preprocessor include certain code segments. This can be very useful
for debugging purposes. For example, in a program 

\begin{verbatim}
 #if (DEBUG == 1)
     printf("DEBUG: energy = %f\n",energy[i]);
     assert(energy[i] < 0);
 #endif
\end{verbatim}

\noindent The behavior of the debugging code can then be controlled
by setting the necessary variable in the header of the program:

\begin{verbatim}
 #define DEBUG 1
\end{verbatim}

\noindent Whenever \texttt{DEBUG} is \texttt{1} the debugging
information is compiled into the code and executed, otherwise not. This
allows to track the behavior of certain variables where potential
values are known from scientific insights. For example, in the C code
above, if the energy is positive [the test is performed with the
built-in \texttt{assert( )} C function; for C++ use \texttt{cassert(
)}] the program aborts abnormally.

Once the program seems to produce reasonable results, one should
{\em always} perform {\em science consistency checks} and, if
possible, compare against known data. Note that the latter is not
always possible. Let us assume we are simulating a two-dimensional
ferromagnetic Ising model \cite{yeomans:92}
\[
 {\mathcal H} = \sum_{\langle ij\rangle}J_{ij}S_i S_j,
\;\;\;\;\;\;\;\; S_i \in \{\pm 1\},
\]
on a square lattice with nearest-neighbor interactions $J_{ij} = -1$,
using Monte Carlo methods. A simple back-of-the-envelope
calculation show that the {\em energy per spin} should converge to $-2$
for temperature $T \to 0$. This follows from the fact that the energy
is minimized when all the spins point in the same direction. In
that case, $E = -2N$, where $N$ is the number of spins.  It is
straightforward to simulate the model at close-to-zero temperature for
a tiny system to verify if the result for the energy per spin is $-2$.
If this is the case, it is likely that there is still a bug hidden
somewhere in the program, but the core routines are likely bug free.
Another option is to reduce the program to study a problem with known
results. For example, if a two-dimensional spin-glass is being studied,
changing the random interactions between the spins ($J_{ij} \in\{\pm
1\}$) to a ferromagnetic interaction ($J_{ij} = -1$) allows for a
comparison with known results for the two-dimensional Ising model.

\subsection{Program timing and profiling}

Once the program is bug-free, it is time to ensure that it runs as
fast as possible. For example, the simulation of spin glasses at
finite temperatures \cite{binder:86,mezard:87,diep:05} usually takes
a large amount of time. A typical project requires $10^5$ or more
CPU hours that corresponds to approximately 12 years. If we can
ensure that our program runs 10 times faster, this would mean that
execution only takes 1.2 years. Therefore {\em profiling} programs
to find parts which are executing slowly is of great importance.

\paragraph{time} The simplest way to time a computer program is to
use the *nix built-in \texttt{time} command:

\begin{verbatim}
 > time ./runme
\end{verbatim}

\noindent The syntax of the output depends on the *nix flavor used. In
all cases the following three numbers are returned: \texttt{real} time,
which is the time it took the process to complete, \texttt{user} time,
which is the time the process spent in user space, and \texttt{system}
time, which is the time the process spent performing system calls.

One can also use the C function \texttt{time( )} in the
program to perform one's own timing. This is done by including
``\texttt{sys/times.h}'' in the header of the program. While this
results in total running times of the program and, if properly implemented,
time step measurements, it does not return information
on how much time was spent on each individual subroutine.

\paragraph{gprof} The GNU graph profiler \texttt{gprof}
\cite{loukides:97} provides a more detailed analysis of the
program. For a program to be profiled, it needs to be compiled with
the ``\texttt{-pg}'' flag:

\begin{verbatim}
 > gcc -pg -o runme.prof main.c lala.c
\end{verbatim}

\noindent Once the program has been compiled, it can be run in the
usual way:

\begin{verbatim}
 > ./runme.prof 
\end{verbatim}

\noindent In addition to the usual output, a file \texttt{gmon.out}
is produced. This file contains the necessary profiling information. At
this point, the profiler is invoked with

\begin{verbatim}
 > gprof runme.prof gmon.out >& output.txt
\end{verbatim}

\noindent Note that the output is delivered to the standard output
and contains a wealth of information, which is why we pipe it into a
file \texttt{output.txt}. The most relevant information is the table
at the end which shows the (flat) profile listing all functions and
how much relative time was spent on them (first column):

\begin{verbatim}
 %     cumulative   self              self     total           
 time   seconds   seconds    calls  ms/call  ms/call  name    
 17.7       3.72     3.72 13786208     0.00     0.00  Append [8]
  6.1       5.00     1.28   107276     0.01     0.03  MkPath [10]
  2.9       5.60     0.60  1555972     0.00     0.00  StringFree [35]
\end{verbatim}

\noindent The second column contains the cumulative time added up
from the time each routine used in the third column. The fourth
column lists how many times a routine was called. The fifth column
contains the average number of seconds per call to a function, i.e.,
the third column divided by the fourth. The final column has the
total time spent on a function and its descendants.  The output also
contains a {\em call graph} that shows all functions and dependencies
(not discussed here, see the man page). In general, it is easy to
spot where the simulation is spending most of the execution time. If
this is not in the core of the program, then there is likely a slow
or cumbersome implementation of a function which needs to be optimized.

\paragraph{gcov} GNU's \texttt{gcov} is a test coverage program
\cite{comment:gcov} that describes the degree to which the source code
of a program has been tested. It therefore allows you to discover
which parts of your program have not been tested. While gprof tells
you how much time is spent in a certain part of the code, \texttt{gcov}
tells you which lines of the code are actually executed and how often.
Using \texttt{gcov} is simple: Compile your program in the following way

\begin{verbatim}
 > gcc -fprofile-arcs -ftest-coverage main.c
\end{verbatim}

\noindent Execute the program (this will generate additional files
needed by \texttt{gcov}) and then run \texttt{gcov} on the individual
c files of your routines:

\begin{verbatim}
 > gcov main.c
\end{verbatim}

\noindent The last step generates a file \texttt{main.c.gcov}
which contains the necessary information to further optimize your
program. In particular, it contains the number of times a function,
statement, or line of the code has been executed.

\section{Running the code}
\label{sec:running}

Before the large production runs are started, it is important to
check how many resources the project will require. For example,
if for a large system the code runs longer than what is allowed
on a supercomputer's queue, the program will be terminated before
completion. Similarly, if the memory requirements exceed the RAM on
the computer, the machine will crash. And if the output exceeds the
free disk space on the computer, a crash will likely occur as well.
A simple approach is to perform {\em scaling checks} where certain
relevant parameters are increased monotonically to see how running
time and memory requirements scale.  A simple example is the number
of particles $N$ in a n-body simulation.  By calculating the running
time $\tau$ for several small and manageable systems one can estimate
$\tau \sim f(N)$ with $f(N)$ a function of the size of the input.
If configurational averages are required---e.g., when performing
disorder averages for spin glasses---these need to be factored into
the total running time as well.

It is also useful to have the program produce periodic output steps
during run time thus providing information on the progress of the
simulation [e.g., via the built-in \texttt{time( )} function].
This has the advantage that further planning is possible during
run time and the user has an idea on how much longer a simulation
could take. A possible (machine-dependent output) to a logfile could
be the header of the data files shown in Sec.~\ref{sec:provenance},
as well as the following useful information:

\begin{verbatim}
# process ID  :      7772
# hostname    :      moo.tamu.edu
# job start   :      Fri Apr 10 18:26:24 2009

# 2    samples:      50  % done
# elapsed time:      0 : 00 : 11.5
# timestamp   :      Fri Apr 10 18:26:37 2009

# 4    samples:      100 % done
# elapsed time:      0 : 00 : 23.1
# timestamp   :      Fri Apr 10 18:26:50 2009

# job end     :      Fri Apr 10 18:26:50 2009
\end{verbatim}

\noindent The logged data show the process ID (PID) of the process
in case it must be killed or paused, the hostname of the machine it
is running on, as well as time-stamped progress reports on the data
generation. Once the program exits, a final time stamp is printed.
With this information, an arbitrary number of simulations on a large
set of workstations can be managed easily, especially if they use a
shared file system.

\subsection{Managing simulations on large numbers of workstations}

Some projects are {\em embarrassingly} parallel (this means that
several simulations can be run in parallel without having them ``talk''
to each other) and thus can be simulated using workstation farms.
Typical problems that fall into this category require configurational
averages, i.e., the simulations have to be repeated many thousand times
with either different parameters, data sets, or initial conditions.
In general, it is recommended to use ready-set software such as
Condor \cite{comment:condor} which simplifies this task considerably
by managing job submission and migration in case of a node failure.
Most universities have at least one condor pool in an effort to tap
into wasted cycles of desktop computers.  The drawback of Condor
is that only very specific compilers can be used and thus some
codes either cannot be compiled at all or might not run at optimal
speeds.  Furthermore, the administrator configuration of condor can
be cumbersome.  By combining secure shell (\texttt{ssh} via OpenSSH)
with some shell scripting, it is straightforward to obtain similar
functionality.

Secure shell is a program for logging into a remote machine and for
executing commands on a remote machine.  It is intended to replace
rlogin and rsh, and provide secure encrypted communications between
two untrusted hosts over an insecure network.  X11 connections and
arbitrary TCP ports can also be forwarded over the secure channel.
Furthermore, by generating a keypair with \texttt{ssh-keygen}
password-less logins can be performed securely. The reader is referred
to the ssh manpages which contain a wealth of information. In this
context only the basic functionality of ssh is presented within the
context of managing several simulations on different workstations.
It is assumed that the user has set up a ssh key pair to allow for
password-less secure login.

Suppose that we have 3 computers, \texttt{moo}, \texttt{boo} and
\texttt{goo} and we want to start simulations on them, manage the
jobs, and possibly kill or pause them. A simple shell script wrapper
(called \texttt{doo.sh}) which can be modified to accommodate all these
operations is the following

\SourceCodeLines{99}
\begin{Verbatim}[fontsize=\small]
 #!/bin/sh

 list="moo.tamu.edu boo.ethz.ch goo.ehtz.ch"

 for mach in $list ;
 do
     echo "$mach:"
     ssh $mach -n $1 $2 $3 $4 $5
     echo "================================================="
 done
\end{Verbatim}

\noindent The first line calls the shell; in the third line we have
a list of all machines we want to manage followed by a loop (lines 6
-- 10) which calls the function \texttt{ssh \$mach -n \$1 ...}. In
the \texttt{ssh} command the option \texttt{-n} redirects standard
input from \texttt{/dev/null}. The different \texttt{\$}i take the
options from the command line and pass them to the ssh command. In
the following example we check the UTC date on the machines with
the command \texttt{date -u}:

\begin{verbatim}
 > ./doo.sh date -u
 moo.tamu.edu:
 Fri Apr 10 23:56:21 UTC 2009
 ===================================================
 boo.ethz.ch:
 Fri Apr 10 23:56:24 UTC 2009
 ===================================================
 goo.ehtz.ch:
 Fri Apr 10 23:56:28 UTC 2009
 ===================================================
\end{verbatim}

\noindent Therefore, any set of instructions can be distributed to a
list of machines remotely. Clearly, using a shell script is rather
rudimentary, and with the use of Perl, much fancier distribution
methods can be accomplished. To start some jobs (\texttt{runme})
on the target machines in a directory called \texttt{run}, replace
line 8 in \texttt{doo.sh} (\texttt{ssh \$mach -n \$1 \$2\ldots}) with

\SourceCodeLines{9}
\begin{Verbatim}[fontsize=\small]
     ssh $mach -n "cd run; ./runme >& logfile &"
\end{Verbatim}

\noindent This changes the working directory to the directory
\texttt{run} and executes the command \texttt{./runme >\& logfile
\&}. To kill the jobs on all machines, use the original version of
\texttt{doo.sh} in combination with Unix \texttt{killall} command:

\begin{verbatim}
 > ./doo.sh killall runme
\end{verbatim}

\noindent Finally, to check on the status of the (running) simulations,
replace line 8 in \texttt{doo.sh} with

\SourceCodeLines{9}
\begin{Verbatim}[fontsize=\small]
     ssh $mach -n "ps axo user,pid,%cpu,%mem,nice,stat,time,cmd \
                   \| egrep \"R \|RN \| RN+ \| R+ \""
\end{Verbatim}

\noindent Note that the command should be placed on one line and
depends somewhat on the *nix flavor. Some testing with the \texttt{ps}
command might be required. When run, it returns the status of the
{\em running} programs only:

\newpage

\begin{verbatim}
 > ./doo.sh 
 moo.tamu.edu:
 USER       PID %CPU %MEM  NI STAT     TIME CMD
 hgk      19995 97.3  4.0  18 RN   20:43:59 ./runme
 ===================================================
 boo.ethz.ch:
 USER       PID %CPU %MEM  NI STAT     TIME CMD
 hgk      32345 93.3  4.0  18 RN   20:32:49 ./runme
 ===================================================
 goo.ehtz.ch:
 USER       PID %CPU %MEM  NI STAT     TIME CMD
 hgk      13455 95.3  4.0  18 RN   20:41:29 ./runme
 ===================================================
\end{verbatim}

\noindent These are some simple tools that make managing jobs on up
to approximately 100 workstations relatively straightforward.

\subsection{HPC clusters}

Explaining the details on how to run simulations on high-performance
computer (HPC or supercomputer) clusters goes far beyond the scope
of this lecture. Furthermore, there are many different ways to set up
computer clusters and many different job schedulers (LSF, LoadLeveler,
PBS, \ldots). The latter are all similar, but unfortunately, the
command line syntax is always different. There are, however, some
basic steps that need to be performed (compilation and job submission)
which are henceforth outlined. Probably the most important aspect
to keep in mind is that programs should be thoroughly tested before
being run on a cluster.  First, wasting CPU time could be very costly,
second, crashing the cluster might signify termination of the account.

Most clusters nowadays use \texttt{modules}. Because applications and
compilers require several environmental variables, modules have been
introduced. The idea is to call the necessary software via a module thus
ensuring that the system configuration is properly set. For example, 
to check on the available modules and then load the Intel compiler type:

\begin{verbatim}
 > module avail
 > module load intel
\end{verbatim}

\noindent After these commands have been issued, compilation with
\texttt{icc} can be accomplished. The command \texttt{module help}
lists all possible options. In general, jobs cannot be simply
started, they need to be sent trough the {\em scheduler}. Using LSF,
a bare-bones job submission can be accomplished via

\begin{verbatim}
 > bsub -n nprocs -W hh:mm ./runme
\end{verbatim}

\noindent Here \texttt{nprocs} is the number of requested
processors and \texttt{hh:mm} represents the wall-clock time limit
of the simulation in hours:minutes.  Note: Only in {\em very}
few very well managed clusters is this task as easy. In general,
many more steps are required, that are thoroughly described in
the documentation of the used scheduler. In the case of parallel
simulations, compilation against MPI libraries \cite{comment:mpi}
(or OpenMP) is required. Many supercomputer centers offer free online
tutorials on code parallelization.

\section{Libraries}
\label{sec:libraries}

Libraries are collections of routines and useful data types to be used
in other programs. The idea is to have well-tested routines which are
optimally programmed available for application development without
having to program them from scratch. For example, we take for granted
to use the \texttt{sin( )} function which is actually part of the math
library \texttt{libm.a}. Therefore, when we use \texttt{sin( )} in a
program, during compilation we need to link to the math library via
\texttt{-lm}.  In this section a quick Howto to generate libraries is
given, followed by useful library packages for scientific computation.

\subsection{Building libraries}

Creating a library is a straightforward task: Suppose you have some
programs that are generic and can be used by many of your applications,
in this example \texttt{myvecprod.c}, \texttt{mydate.c} and
\texttt{myanalyze.c}, together with a header file \texttt{mytools.h}
which contains the necessary data types and function prototypes. The
goal is to generate a library \texttt{libmytools.a}. First, object
files need to be created:

\begin{verbatim}
 > gcc -O2 -c myvecprod.c mydate.c myanalyze.c
\end{verbatim}

\noindent Keep in mind that if you do not optimize the code on
compilation the library calls will be slow! Therefore, as a reminder,
in the previous example a \texttt{-O2} flag has been added. The
library is created with the Unix command \texttt{ar}

\begin{verbatim}
 > ar r libmytools.a myvecprod.c mydate.c myanalyze.c
\end{verbatim}

\noindent For further options of \texttt{ar} please look at the man
page. After including new object files you have to update the internal
object table of the library with

\begin{verbatim}
 > ar s libmytools.a
\end{verbatim}

\noindent Once the library has been created, suppose it is
placed in \texttt{/home/user/mytools/lib/} and the header in
\texttt{/home/user/mytools/include}. To compile a program using
routines from the header files and library, simply do

\begin{verbatim}
 > gcc -o runme myprog.c -I/home/user/mytools/include \
        -L/home/user/mytools/lib/ -lmytools
\end{verbatim}

\noindent By default, libraries are {\em dynamically} linked. This
means that the code of the library is not included in your program
and only called on run-time. This can be a problem if you use the
program on other machines where the libraries are not installed. To
overcome this problem {\em static} linking can be performed with the
flag \texttt{-static} (or similar, see the compiler documentation)
at the price of a larger executable.

\subsection{Built-in C/C++ libraries}

\paragraph{Standard C library} The standard C library
\cite{comment:stdlib} is included by default on compilation. The
only requirement is to include the header file \texttt{\#include
<stdlib.h>}. Most of the core operations such as memory allocation
are included in this library. Therefore, its functions are very
well documented in the man pages.  For example, \texttt{man malloc}
returns a detailed description of memory allocation functions,
as well as their prototypes. There are many different header files
in the standard library that can be included and thus expand the
functionality of a program \cite{comment:stdlib}. The most useful ones
are

\begin{itemize}

\item[]{\hspace*{-1em}\texttt{stdio.h}\newline Handles input/output
of the program.  }

\item[]{\hspace*{-1em}\texttt{math.h}\newline Contains a multitude
of useful mathematical functions (linking to the math libraries is
needed via \texttt{-lm}).}

\item[]{\hspace*{-1em}\texttt{stdlib.h}\newline Contains core functions
for memory allocation, type conversion operators, (bad) random number
generators, as well as routines for process control and interaction
with the environment (e.g., system calls).}

\item[]{\hspace*{-1em}\texttt{time.h}\newline Declares time and date
functions that provide standardized access to time/date manipulation
and formatting.  }

\end{itemize}

\paragraph{Standard template library (STL)} The STL is part of the C++
Standard Library. It provides {\em containers} to store objects, {\em
iterators} for container access, {\em algorithms}, and {\em functors}
(function objects) \cite{stroustrup:00}. It is far beyond the scope
of the lecture to discuss the STL, the reader is referred to the
documentation freely available on SGI's website \cite{comment:stl}.

\subsection{Scientific libraries}

\paragraph{GNU Scientific Library} The GSL is a vast collection of
routines aimed at solving numerical problems.  It is freely available
\cite{comment:gsl} and contains routines for the following numerical
tasks:

\begin{itemize}

\item[$\Box$]{Complex number arithmetic}

\item[$\Box$]{Polynomials and root finding}

\item[$\Box$]{A vast list of special functions \cite{abramowitz:64}}

\item[$\Box$]{Vector and matrix operations}

\item[$\Box$]{Interpolation, differentiation, integration}

\item[$\Box$]{Permutations and sorting}

\item[$\Box$]{Random number generation}

\item[$\Box$]{Statistical analysis}

\item[$\Box$]{Monte Carlo methods}

\item[$\Box$]{Fast Fourier transforms, etc.}

\end{itemize}

\noindent Furthermore, the GSL interfaces with the BLAS (basic linear
algebra subroutines in FORTRAN) which are the standard in linear
algebra operations.  To use the GSL, the necessary header files need
to be included in the program.  In the following example, the GSL is
used to generate uniform random numbers:

\SourceCodeLines{99}
\begin{Verbatim}[fontsize=\small]
 #include <stdio.h>
 #include <gsl/gsl_rng.h>
 
 int main()
 {
     gsl_rng *rng;                             /* pointer to RNG */
     int    i;                                       /* iterator */ 
     int    n = 10;                  /* number of random numbers */
     double u;                                  /* random number */

     rng = gsl_rng_alloc(gsl_rng_mt19937); /* allocate generator */
     gsl_rng_set(rng,1234)                 /* seed the generator */
     
     for(i = 0; i < n; i++){
         u = gsl_rng_uniform(rng);    /* generate random numbers */
         printf("%f\n", u);
     }
     
     gsl_rng_free(rng);                      /* delete generator */
     
     return(0);
 }
\end{Verbatim}

\noindent The numbers depend on the seed used by the generator. First
the generator is allocated with \texttt{rng = gsl\_rng\_alloc(}{\em
generator}\texttt{)} and seeded with \texttt{gsl\_rng\_set(rng,}\linebreak{\em
seed}\texttt{)}.  Finally, the uniform random numbers in the interval
$[0,1[$ are produced with\newline\texttt{gsl\_rng\_uniform(rng)}. Using
the pre-defined GSL functions saves hours of programming and testing.
The documentation of the GSL is extensive and usually has example
problems in each section.

\paragraph{Numerical Recipes}

The Numerical Recipes \cite{press:95} (NR) is a large collection
of routines for many numerical problems (similar to the GSL)
accompanied by a very useful book. The Numerical Recipes have been
ported to different programming languages, such as C, C++, FORTRAN,
FORTRAN90 and Pascal. The book used to be accessible online for free,
but unfortunately this seems to have changed recently. The advantage
over the GSL is that the routines have been ported to many programming
languages, but the coding style is slightly outdated. Still, it
is always a good idea to look at the NR book before you start any
numerical project. Note that the single-user license allows (in theory)
for only {\em one} executable to be run at a given point in time. Thus,
to run software on multiple CPUs a group license is recommended.

\paragraph{Boost libraries} The Boost libraries \cite{comment:boost}
provide free peer-reviewed portable C++ source libraries that
extend the C++ STL.  In order to ensure efficiency and flexibility,
Boost makes extensive use of {\em templates}. Boost has been a
source of extensive work and research into generic programming
and metaprogramming in C++.  The current Boost release contains
approximately 80 individual libraries for a variety of tasks, such 
as

\begin{itemize}

\item[$\Box$]{Linear algebra}

\item[$\Box$]{Random number generation}

\item[$\Box$]{Regular expressions}

\item[$\Box$]{Graphs}

\item[$\Box$]{Math special functions and statistics}

\end{itemize}

\paragraph{LEDA libraries}

LEDA \cite{mehlhorn:99} is a library dedicated to the efficient
implementation of data types and algorithms. The library is
commercial, although there is also a (crippled) free version
\cite{comment:leda}. Although LEDA is written in C++, it can also be
accessed from standard C code. The free version includes the following
data types:

\begin{itemize}

\item[$\Box$]{Data types for strings and multidimensional arrays}

\item[$\Box$]{Number data types (arbitrary precision)}

\item[$\Box$]{Stacks and queues, sets and trees}

\item[$\Box$]{Graphs (directed and undirected, labeled)}

\item[$\Box$]{Dictionaries (similar to Perl's hash)}

\item[$\Box$]{Data types for two- and three-dimensional geometries
(points, spheres, etc.)}

\end{itemize}

\noindent Furthermore, all necessary operations to manipulate standard
data types are included. Finally, the library contains routines to
compute strongly-connected components, shortest paths, maximum flows,
minimum-cost flows, and minimum matchings for graphs, that are very
useful when studying complex systems in statistical mechanics.

\section{Data reduction and analysis with Perl}
\label{sec:reduction}

Perl is a high-level, general-purpose, interpreted, dynamic programming
language.  Perl is similar to C in syntax, but also borrows elements
from shell scripting, awk and sed.  Perl was originally developed
to make report processing easier.  Therefore, the language provides
powerful text processing facilities without the arbitrary data length
limits of many contemporary Unix tools, facilitating easy manipulation
of text files. Because of its flexibility, Perl has been nicknamed
``the Swiss Army chainsaw of programming languages.'' O'Reilly Media
Inc.~has published many books \cite{comment:oreilly} about Perl,
all of which are very useful.

It would be impossible to present the vast capabilities of Perl in
this short lecture, therefore it is recommended that the reader looks at
the literature \cite{comment:oreilly}.  To illustrate the use and
simplicity of Perl, let us construct a few simple examples. Because
Perl is a scripting language (although it can also be compiled), the
typical syntax for a Perl script (called here \texttt{myscript.pl}) is

\SourceCodeLines{9}
\begin{Verbatim}[fontsize=\small]
 #!/usr/bin/perl -w 
 # this is a comment

 print("hello world!\n");
\end{Verbatim}

\noindent On the command line, this script can then be executed with
the Perl interpreter

\begin{verbatim}
 > perl myscript.pl
 hello world!
\end{verbatim}

\noindent The Perl syntax is very flexible, for example,
\texttt{print("hello world!\textbackslash n");} could be replaced
by \texttt{print "hello world!\textbackslash n";}. Each statement is
terminated by a semicolon, similar to other high-level languages.
Suppose we have a set of {\em several} files which contain tabulated
numbers (see Sec.~\ref{sec:provenance} for an example) and we want
to compute an average over the third column in the files. Here is a
Perl script to accomplish this task (called \texttt{average.pl}):

\SourceCodeLines{99}
\begin{Verbatim}[fontsize=\small]
 #!/usr/bin/perl -w 

 if($#ARGV < 0){
     warn "\n\toops! type $0 files.\n"; exit; 
 }
 else{
     @files = @ARGV; 
 }

 @table   = ();            # array where data are stored
 $counter = 0;             # counter 

 foreach $file (@files){   # loop over files
     open(DATA, "$file");  # open file
     while (<DATA>){       # parse file
         chomp;            # remove newlines 
         if ( $_ =~ /^#/){ # skip comment lines in files 
             next;
         }           
         $table[$counter] = (split /\s+/)[2];
         $counter++;
     }
     close(DATA);
 }

 $mean = 0;
 for($i = 0; $i < $counter; $i++){
     $mean += $table[$i];
 }
 $mean /= $counter;

 printf("mean = %4.3f\n",$mean);
\end{Verbatim}

\noindent Running the perl script on multiple data files on the
command line one obtains:

\begin{verbatim}
 > perl average.pl *.dat
 77.72
\end{verbatim}

\noindent Let us dissect \texttt{average.pl} and highlight the
advantages of using the Perl programming language.  In line 1 the Perl
interpreter is called. The option \texttt{-w} turns on warnings which,
in general, is useful to do when debugging.  From lines 3 -- 8 we
first check if files have been given as arguments to the script. If
not, we exit with an error message. If yes, the file names are
stored in an array \texttt{@files}. In Perl, arrays always start with
``\texttt{@}.'' Variables usually start with a \texttt{\$}, such as in
the initialization of \texttt{\$counter}. The array \texttt{@table}
is initialized to be empty.  One useful function in Perl is the
\texttt{foreach( )} function. In this case it allows us to loop over
all file names in \texttt{@files} and manipulate them.  A data stream
is opened in line 14. While the data are being parsed line by line,
newlines at the end of the lines are removed with \texttt{chomp}
in line 16. Line 17 highlights the first major strength of Perl:
full support of regular expressions. We check that the currently
parsed line (\texttt\$\_) does not start with a hash by matching the
string \texttt{/\^{}\#/} using the matching operator \texttt{\~}. If
it does, we skip the line (\texttt{next}). In line 20 the data on
each line of the file are split by (regexp) matching white spaces and
we select the number in the third column (indices start with zero)
to be stored in \texttt{\$table[\$counter]}. Once the file has been
parsed, the data stream is closed. After all files in \texttt{@files}
have been processed, we compute---in this simple example---the mean
of the numbers.

Clearly, there are far more elegant ways of performing this task
in Perl. In an attempt to keep the syntax simple, readable, and as
close as possible to C, we have written the program in this way.
Furthermore, performing an average is a trivial task that could be
performed on the command line with the \texttt{cat} and \texttt{awk}
commands. Far more complex data operations can be performed
once the data are read into memory.  Further features of Perl:

\begin{itemize}

\item[$\Box$]{In addition to the variable and array data types, Perl
also allows {\em hashes} which can be seen as {\em generalized} arrays
where the {\em indices} can be arbitrary quantities and do not have
to be numbers. Example: \texttt{\%kitchen = (apple => 'red', banana =>
'yellow')}. Calling \texttt{\$kitchen\{apple\}} returns \texttt{red}.}

\item[$\Box$]{Support for subroutines.}

\item[$\Box$]{Full support for regular expression pattern matching.}

\item[$\Box$]{Perl {\em modules} from CPAN \cite{comment:cpan}. These
are similar to libraries in C. CPAN provides a huge list of modules
for almost any thinkable task. For example, interfaces to eBay, image
manipulation, server and demon utilities, Boost data types,
to name a few. These can be installed and then called from Perl
scripts.}

\item[$\Box$]{Full portability of the code. Since the program is
interpreted on the fly and does not require compilation, Perl scripts
can be run on many different platforms and architectures, as long as
the interpreter is installed.}

\end{itemize}

\noindent Although Perl is not very fast, the ease of use (especially
for C programmers) makes it the language of choice for data post
processing.

\section{Data visualization and fitting: gnuplot}
\label{sec:viz}

Once the simulation data have been produced and reduced, analysis
and visualization need to be performed. There are many software
packages that accomplish these tasks, each with their strengths
and weaknesses. For example, data plotting can be accomplished with
xmgrace (GUI plotting application), gnuplot (command line), SuperMongo
(scripting language) or Mathematica. One (freely-available) software
package which has good data fitting capabilities and can produce
decent-looking figures is gnuplot \cite{comment:gnuplot}. In this
section the core functionality of gnuplot is presented.  Gnuplot can
be started on any *nix system via

\begin{verbatim}
 > gnuplot
\end{verbatim}

\noindent Note that gnuplot can also be used in batch mode
(running scripts without direct users interaction). This is useful for
interfacing with data analysis tools, as well as plotting many images.
Gnuplot has an extensive built-in help system, which can be called via
the \texttt{help} command. \texttt{help help} gives an introduction
to the built-in help.

\subsection{Plotting data}

Gnuplot can be used both to do quick-and-dirty plots of data to obtain
an impression on how things look, as well as publication-quality
figures.

\begin{figure}[!htb]
\centering
\includegraphics[scale=0.50]{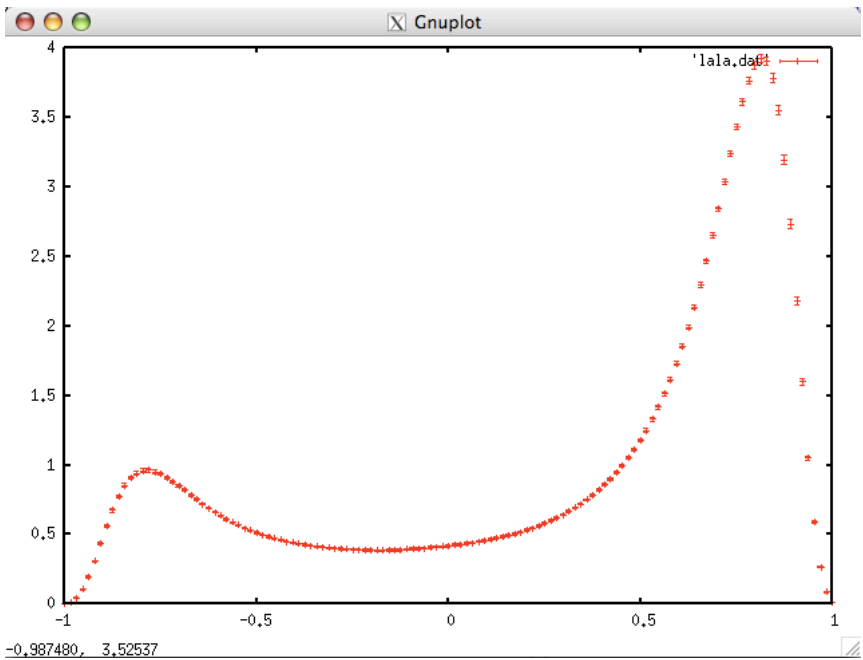}
\caption
{
Sample X11 terminal output from gnuplot for a quick-and-dirty plot.
}
\label{fig:gpu}
\end{figure}

\paragraph{Quick-and-dirty} In general, for quickly displaying
tabulated data of the form \texttt{x~y~err\_y} stored in a file
\texttt{results.dat} the following commands have to be issued at the
gnuplot prompt:

\begin{verbatim}
 gnuplot> plot [0:3][-2:10] "results.dat" with yerrorbars
\end{verbatim}

\noindent The previous command plots the numbers in
\texttt{results.dat} in the \texttt{x}-range $[0,3]$ and the
\texttt{y}-range $[-2,10]$ with error bars. Note that `\texttt{with
yerrorbars}' can also be abbreviated with `\texttt{w e}.' A X11 window
pops up displaying the data (see Fig.~\ref{fig:gpu}).  Placing
the cursor {\em over} the X11 window with the data and hitting the
\texttt{h}-key lists all possible {\em interactive} commands
in the terminal window. For example, hitting the \texttt{L}-key
close to an axis near the cursor changes the axis from linear
to logarithmic. Finally, to see all plotting options as far as line
styles and symbols is concerned, simply enter the command \texttt{test}
in the command line.

\bigskip

\bigskip
\bigskip

\paragraph{Publication-quality plots} To generate ``presentable''
plots of data, it is highly recommended to write a gnuplot script
(\texttt{macro.gp}) and generate an encapsulated PostScript file in
batch mode:

\begin{verbatim}
 > gnuplot macro.gp
\end{verbatim}

\noindent The reason is that many options need to be set and the
figure tweaked to ensure that it is nice and fits the journal format.
The following script produces publication-quality images. Note
that these can and must be further improved. This is merely a
proof-of-concept macro:

\SourceCodeLines{99}
\begin{Verbatim}[fontsize=\small]
 # preamble (set parameters)
 set size ratio 1 2,2
 set terminal postscript color eps enhanced "Times-Roman" 48
 set output "results.eps"
 set origin 0,0
 unset label
 set size square
 set pointsize 1.5
 set key 0.3, 3.5

 # x-axis tick, ranges and labels
 set xrange [-1.0:1.0]
 set xtics -1.0,0.5,1.00
 set mxtics 5
 set xlabel "{/Symbol l}"

 # y-axis ticks, ranges and labels
 set yrange [0.0:4.0]
 set ytics 0,1,4
 set mytics 5
 set ylabel "{/Times-Italic P}_{c}({/Symbol l})"

 # add other text labels
 set label "(b)" at -0.75, 3.50
 set label "{/Times-Italic T} = 0.20" at 0.4, 0.5

 # plot points with errors, draw line trough them
 set multiplot
 plot  "results.dat" using 1:2:3 with yerrorbars pointtype 5 \
         ti "{/Times-Italic N} = 128"
 plot  "results.dat" using 1:2 with lines notitle lt -1
 unset multiplot

 # fix bounding box with perl
 !perl -pi -e 's/%%BoundingBox: 50 50 770 554/%%BoundingBox: 50 50 560 560/'\
         results.eps
\end{Verbatim}

\begin{SCfigure}[][!htb]
  \includegraphics[scale=0.35]{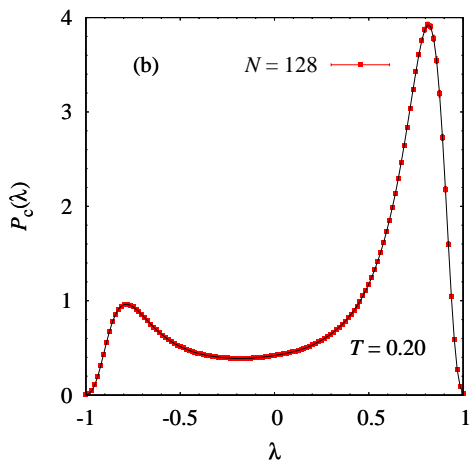}\hspace{2pc}
  \caption
   {
    Print-quality figure generated with the presented gnuplot macro.
    It shows the same data as in Fig.~\ref{fig:gpu}. The aspect ratio
    has been set to unity to better fit a two-column format. This can
    be undone by removing the line `\texttt{set size square}' in the macro.
    \vspace{4pc}
   }
  \label{fig:gpn}
\end{SCfigure}

\noindent The most relevant lines and commands are explained
below. For further information on the different commands used,
the reader is referred to the gnuplot documentation or the
Howto by T.~Kawano \cite{comment:kawano}.  Line 3 determines the
\texttt{terminal} to use. Default is X11 (display to the screen),
in this case we change it to produce an encapsulated PostScript file
with a \texttt{Times-Roman} font of 48 points and the file name
``\texttt{results.eps}'' (line 4). The blocks starting at lines 11
and 17 determine the look and labels of the axes. Note that specific
fonts can be called to render the labels, including the Greek alphabet
(see Ref.~\cite{comment:kawano} for details).  Further text labels
are added after line 23. In line 28 gnuplot is instructed to plot
multiple data sets. The reason is that in line 29 we plot the data
points as red squares with error bars, and in line 31 we plot the
same data again connecting the points, but without symbols.  Finally,
in lines 35/36 we use a system call to the shell invoking perl to
change the bounding box of the resulting encapsulated PostScript such
that the figure is square (better for two-column formatted journals).
The output is shown in Fig.~\ref{fig:gpn} and is clearly superior
than the plot shown in Fig.~\ref{fig:gpu}.

\subsection{Fitting data}

Gnuplot also offers many options for data manipulation. A very
useful one is data fitting. Suppose we want to fit some data
file of the form \texttt{x~y~err\_y} to an exponential decay of
the form $f(x) = a\exp(-bx)$ with $a$ and $b$ parameters.  At the
\texttt{gnuplot} prompt we first enter the function followed by some
starting values. The starting values are ``good guesses'' of what
the parameters should be to help the Levenberg-Marquardt nonlinear
least-squares fitting routine converge.

\begin{verbatim}
 gnuplot> f(x) = a*exp(-b*x)
 gnuplot> a = 100.0 
 gnuplot> b = 2.0
\end{verbatim}

\noindent The fit to data in a file \texttt{results.dat} is then
performed using the \texttt{fit} function

\begin{verbatim}
 gnuplot> fit f(x) "results.dat" using 1:2:3 via a,b
\end{verbatim}

\noindent The subcommand ``\texttt{using 1:2:3 via a,b}'' tells gnuplot
to use the first, second and third column for \texttt{x~y~err\_y},
respectively and to fit with $a$ and $b$ as parameters. We could
also, for example, fix $a$ and only vary $b$ in the fit by changing
``\texttt{via a,b}'' to ``\texttt{via b}.'' Gnuplot then attempts
to fit the curve to the data and outputs a log of the iterative process,
as well as the optimal parameters (with error bars) and the $\chi^2$
per degree of freedom of the fit \cite{press:95} (measure of the
quality of the fit).  Furthermore, a correlation matrix between the
parameters is provided. Typical (abridged) output:

\begin{figure}[!htb]
\centering
\includegraphics[scale=0.51]{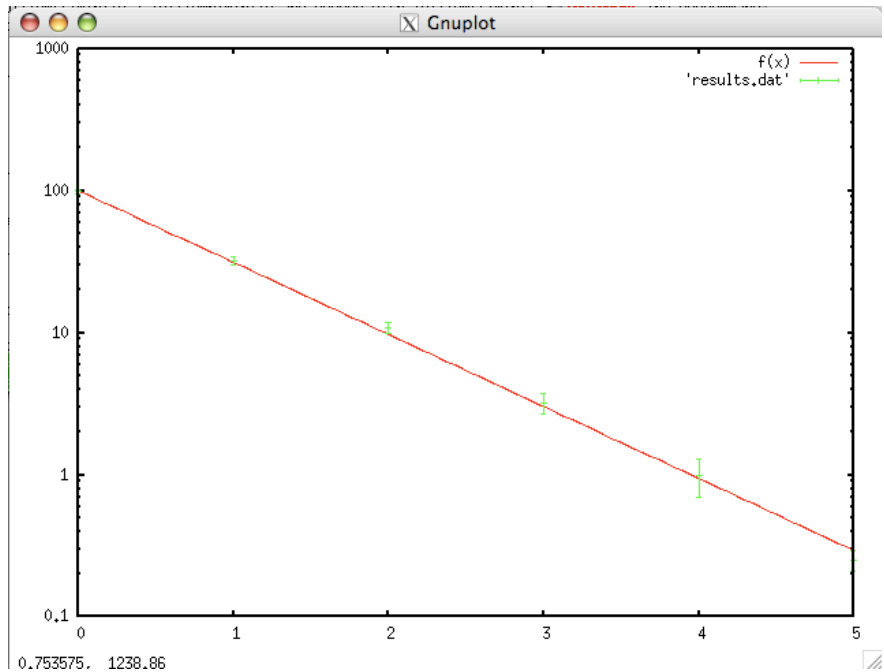}
\caption
{
Visual verification that the fitting routine has converged to the
correct minimum. The fit (solid line) follows the data points
closely.
}
\label{fig:gpf}
\end{figure}

\newpage

\begin{verbatim}
 After 5 iterations the fit converged.
 final sum of squares of residuals : 3.09786
 rel. change during last iteration : -2.86707e-16

 degrees of freedom (ndf) : 4
 rms of residuals      (stdfit) = sqrt(WSSR/ndf)      : 0.880036
 variance of residuals (reduced chisquare) = WSSR/ndf : 0.774464

 Final set of parameters            Asymptotic Standard Error
 =======================            ==========================

 a               = 100.219          +/- 2.486        (2.481%)
 b               = 1.16714          +/- 0.01981      (1.698%)

 correlation matrix of the fit parameters:

                 a      b      
 a               1.000 
 b               0.436  1.000 
\end{verbatim}

\noindent In this case the optimal parameters are $a = 100.2 \pm 2.5$
and $b = 1.167 \pm 0.019$.  One should always also verify the quality
of the fit visually via

\begin{verbatim}
 gnuplot> plot f(x), "results.dat" w e
\end{verbatim}

\noindent In this case, the fitting routine seems to have converged to
the proper minimum, as can be seen in Fig.~\ref{fig:gpf} (linear-log
plot).

If the fit does not seem to agree with the data or it does
not converge, retry the fitting procedure with new starting values.

\section*{Disclaimer \& Acknowledgments}

I would like to thank Juan Carlos Andresen and Ruben Andrist for
critically reading the manuscript, as well as John Chang for numerous
suggestions on the first edition of the lecture notes.

Finally, I would like to add the disclaimer that the presented
approaches are by no means optimal for simulating all possible
scientific problems. They definitely do not scale to large software
projects with many collaborators and they are purely based on the
author's own experiences in simulating physical systems. But they
{\em do} provide a starting point to tackle problems using numerical
methods.

\newpage

\bibliographystyle{plain}  
\bibliography{comments,refs}

\end{document}